\documentstyle[preprint,aps]{revtex}
\tightenlines
\begin{document}
\draft
\title
{\bf Photodisintegration of the Three--Nucleon Systems and their 
Polarizabilities}
\author{Victor D. Efros$^{1,2)}$, Winfried Leidemann$^{3,4)}$, and
Giuseppina Orlandini$^{3,4)}$}
\address{
1) European Centre for Theoretical Nuclear Physics and Related Areas,
Villa Tambosi, \\
I-38050 Villazzano (Trento), Italy\\
2) Russian Research Centre "Kurchatov Institute", Kurchatov Square 1,
123182 Moscow, Russia\\
3) Dipartimento di Fisica, Universit\`a di Trento, 
 I-38050 Povo (Trento), Italy\\
4) Istituto Nazionale di Fisica Nucleare, Gruppo collegato di Trento, Italy
}

\date{\today}
\maketitle
\begin{abstract}
The total photodisintegration cross sections of three--body nuclei are 
calculated with semirealistic NN potentials below pion threshold. Full final 
state interaction with Coulomb force is taken into account via the Lorentz 
integral transform method. The experimental total cross sections are well
described and the sum rule $\sigma_{-1}(^3$H) agrees with elastic electron
scattering data. The calculated $^3$He polarizability is 0.15 fm$^3$.
\end{abstract}

\vfill\eject

The photodisintegration cross sections of the three--nucleon systems are  
important quantities for understanding the physics of few--body systems. 
As reported in Ref. \cite{Faul81} in the past experimental and theoretical 
work have mostly 
concentrated either on the low--energy two--body or on the low--energy
three--body breakup. On the theoretical side several calculations of the 
process have been performed with separable NN potentials \cite{BaP70,GiL76}. 
In this framework quantitatively 
accurate results have been obtained in Ref. \cite{GiL76}. However,
mixed symmetry states
as well as noncentral and Coulomb NN forces have not been considered. The 
results compare rather well with experiment in the peak region, but 
underestimate the experimental data in the threshold region. 
Approximate three-body
breakup calculations have also been performed with low--energy local NN 
potentials in the framework of the hyperspherical approach \cite{Fang78}. The 
only calculation with a realistic NN force has been carried out in Ref. 
\cite{Vost81} for the $^3$H case. It has been a substantial 
progress, but the calculation still has some shortcomings: the used NN force 
does not provide a sufficiently good fit to NN data; the hyperspherical 
expansion employed to describe the interior part of the continuum wave 
functions retains only the three lowest terms and the error in the
truncation has not been fully estimated. 
The results of the two calculations of Refs. 
\cite{Fang78,Vost81} are only in qualitative agreement with the data.  

An important observable connected with the total photodisintegration cross 
section is 
the electric polarizability $\alpha_{E1}$. It has to be mentioned that the 
experimental $\alpha_{E1}$ value of $^3$He determined from elastic scattering 
on $^{208}$Pb \cite{GLK91} is at variance with the value extracted from 
photoabsorption experiments. 
   
In the present work total photodisintegration cross sections of the
three--nucleon systems are calculated accurately with NN forces of a type 
different from separable potentials. The polarizabilities of the 
three--nucleon systems are calculated as well, both directly,
by means of sum--rules techniques, and 
integrating the properly weighted
total photodisintegration cross sections. We use  local even central NN 
potentials that reproduce the low--energy NN properties and provide a 
realistic description of the $^1S_0$ and $^3S_1$ NN phase shifts up to 
the pion threshold. Specifically, 
we use the same NN potential models, 
Malfliet--Tjon  I+III (MT) \cite{MT69}
and Trento (TN) potentials, as in our works on electro-- \cite{ELO97} and
photodisintegration \cite{ELO96} of the $^4$He nucleus. The MT parameters
are taken from Ref. \cite{KaG92}. The description of the NN scattering 
data given by these potentials has been shown in Ref. \cite{ELO96}.
Besides the strong interaction also the Coulomb force is included in our 
nuclear hamiltonian. This leads to differences between the $^3$He and $^3$H 
cross sections, which will also be  discussed in the following.
  
The photodisintegration cross sections are calculated within the
method of the Lorentz integral transform \cite{ELO94}. 
This approach allows the inclusion of 
the full final state interaction without explicit calculation of the continuum
states. The longitudinal $(e,e')$ form factors of the nuclear
two-- \cite{ELO94}, three-- \cite{Sara95} and four--body systems \cite{ELO97} 
have already been calculated with this method. 

Recently we have also studied
in this way the total cross section of the process 
$\gamma + ^4$He $\rightarrow$ X below pion threshold \cite{ELO96}. We 
have predicted a pronounced giant dipole
resonance. Since the rather flat cross sections for the two--body
breakup channels ($^3$H-$p$, $^3$He-$n$) seem to be settled,
the strong peak has to be attributed to an additional cross section
from the not yet measured $(\gamma,np)$ channel. 
Contrary to the four--body system
the peak of the giant resonance is rather well determined in experiment
for the three--body systems. Thus in this case we have the possibility
to get an experimental confirmation of the results obtained with our 
new approach.

It is well known that the total photoabsorption cross section 
can be reliably calculated in the dipole approximation
\begin{equation}
 \sigma_{T}(E_{\gamma})=4\pi^2(e^2/\hbar c)E_{\gamma}R(E_{\gamma})\,, 
\label{eq:r}
\end{equation}
with
\begin{equation}
 R(E_{\gamma})=\int df |\langle\Psi_f|D_z|\Psi_0\rangle|^2\delta(E_f-E_0-
E_{\gamma})\,,  
\end{equation}
where
\begin{equation}
 \vec D = \sum_{i=1}^Z (\vec r_i - \vec R_{cm})\,. 
\end{equation} 
In Eq.(2) $\Psi_0$ is the three--body bound state wave function with  
energy 
$E_0$, and $\Psi_f$ are final state wave functions normalized as 
$\langle\Psi_f|\Psi_{f'}\rangle=\delta(f-f')$.  In the delta function 
in Eq.(2) we 
neglect the very small nuclear recoil energy. We calculate the response 
function $R$ via evaluation and subsequent inversion of its Lorentz integral 
transform 
\begin{equation}
{\cal L}(\sigma=-\sigma_R+i\sigma_I)=\int_{E_{th}}^\infty 
dE_\gamma {R(E_\gamma)\over(E_\gamma-\sigma_R)^2+\sigma_I^2}\,.
\end{equation}
The transform $\cal{L}(\sigma)$ of the response $R$ is found as
\begin{equation} 
{\cal L}(\sigma)=\langle \tilde{\Psi}(\sigma)| \tilde{\Psi}(\sigma)\rangle, 
\end{equation}
$\tilde{\Psi}$ being the localized solution of the Schr\"odinger--like
equation 
\begin{equation}
(\hat{H}-E_0+\sigma)\tilde{\Psi}(\sigma)=Q
\end{equation}
with the source term $Q=D_z\Psi_0$. 

The wave function $\Psi_0$ is the ground state solution for the same 
hamiltonian. The binding energies for both our potential models are 8.7 MeV 
($^3$H) and 8.0 MeV ($^3$He). Compared to the experimental values we have a 
very small overbinding of 0.2 MeV. For the charge radii (with finite size
nucleons) we obtain the  values  $r_{ch}(^3$H)$=1.76 $ fm,  $r_{ch}(^3$He)$=
1.94$ fm, and $r_{ch}(^3$H)$=1.74 $ fm, $r_{ch}(^3$He)$=1.92$ fm for the MT and 
TN potentials, respectively, which are in a good agreement with the 
experimental radii $r_{ch}(^3$H)$=1.76\pm 0.04$ fm  \cite{Ju85},  $r_{ch}
(^3$He)$=1.976\pm 0.015$ fm \cite{Ott85}. We have checked the reliability
of our ground state calculation by comparison  with the MT potential results
of Ref. \cite{FGP80}. For this check we have used the same MT potential 
parameters as given in Ref. \cite{FGP80}. The agreement
for binding energies, radii and mixed symmetry state probabilities is very good.

We solve Eq. (6) for the multipolarity $L=1$, spin $S={1 \over 2}$, and
isospin $T={1 \over 2}$ and ${3 \over 2}$  with 
the help of the correlated hyperspherical expansion and the hyperradial 
expansion of the same form as in Refs. \cite{ELO97,ELO96}. The $K_{max}$ value 
equals to 7.  The convergence of the transform is well attained with this 
$K_{max}$ value and the convergence rate
is similar to that shown in Ref. \cite{ELO96} for the $^4$He case. 
The values of $\sigma_I=20$ MeV and 5 MeV have been employed.
As for $^4$He the inversion has been 
performed both for $\sigma_I=20$ MeV and for 
a combination of the transforms with $\sigma_I=5$ and 20 MeV,
chosen so that the 
former transform gives a predominant contribution to the very steeply rising 
low--energy wing of the response and the latter to its high--energy wing.
While inverting the transform the n-d low--energy behaviour 
$[E_{\gamma}-(E_{\gamma})_{min}]^{3/2}$ has been incorporated into our trial 
response. Of course for $^3$He this is not the correct breakup behaviour
at the very threshold, but already 1 MeV above it the assumed 
low--energy response should be a rather good approximation. This is also 
confirmed by our sum rule checks (see below). 

For $^3$H we can first compare the 
inverse energy weighted sum $\sigma_{-1}=\int_{E_{th}}^\infty 
E_\gamma^{-1}\sigma_T(E_\gamma)dE_\gamma$ with experiment,
using the calculated value of the radius. In fact this sum is known 
to be entirely determined by the triton point proton radius \cite{OCP69}
\begin{equation}
\sigma_{-1}(^3{\mbox H})=\frac{4\pi}{3}\frac{e^2}{\hbar c}
\langle\, r_p^2(^3{\mbox H}) \,\rangle \,. \label{eq:s}
\end{equation}
Eq. (\ref{eq:s}) is exact and follows directly from the sum rule $\sigma_{-1}=
(4\pi/3)(e^2/\hbar c)\langle\Psi_0|D^2|\Psi_0\rangle$ and the fact that for a 
single--proton nucleus $\vec{D}=\vec{r}_p-\vec{R}_{cm}$. Using the relation
$\langle\, r_p^2(^3{\mbox H})\,\rangle = r^2_{ch}(^3$H)$-r^2_{ch}(p)-
2r_{ch}^2(n)$
with the experimental values $r_{ch}(p)=0.862$ fm \cite{Sim80}, 
$r_{ch}^2(n)=-0.117$ fm$^2$ 
\cite{KK73} and $r_{ch}(^3$H) \cite{Ju85} one obtains $\sigma_{-1}(^3$H)=2.485$\pm$0.135 
mb, which agrees with our sum rule values listed in Table 1. We would like to
point out that the neutron charge radius is not negligible in obtaining
the correct value of $\langle\, r_p^2(^3{\mbox H})\,\rangle$ from 
$r^2_{ch}(^3$H)\cite{FGP80b}. 
For $^3$He a sum rule analogous to Eq.(7) involves  
$\langle\, r_n^2(^3{\mbox He}) \,\rangle$. 
Attempts in the literature to predict
$\sigma_{-1}(^3$He) estimating $\langle\, r_n^2(^3{\mbox He}) \,\rangle$ 
by means of the $^3$He experimental
magnetic radius are much more uncertain. In fact no exact relation involving
only experimental observables exists between these two radii.

Now we turn to the comparison of our results with the experimental data for
the total cross section. The total cross sections have been studied
only in Ref. \cite{Faul81}. In that experiment the triton low--energy total 
cross section was measured as 
the sum of two-- and three--body breakups. In the same experiment also the 
$^3$He three--body breakup cross section was determined. Combined with the 
two--body breakup data from Ref. \cite{Tic73} this led to a total cross 
section also for $^3$He. In the energy range between electric dipole peak and
pion threshold there exists only one experiment. It consists in a measurement 
of the two-- and three--body $^3$He photodisintegration \cite{Fet65}.
The results for the total cross section were 
not listed but since two-- and three--body reactions were both measured for the 
same photon energies we could determine the total cross section summing them.

Our results for the low--energy cross sections are shown in Fig. 1
together with the experimental data. It is seen that
there is a good agreement with experiment in the threshold
region. In the peak region, where one has a slight difference between the
two experimental data sets for $^3$He, our results agree with 
those of Ref. \cite{Fet65}. For triton the agreement with Ref. \cite{Faul81}
is somewhat better. We would like to point out that seemingly only the upper 
experimental curve could lead to a fulfillment of the $\sigma_{-1}$ sum rule.
Both our potential models lead to rather similar cross sections. As in the 
$^4$He photodisintegration \cite{ELO97} one finds a somewhat stronger cross 
section up to about 10 MeV above threshold for the MT potential.
In Fig. 2 we show the results up to pion threshold. Again one has
very similar cross sections with MT and TN potentials for both 
nuclei. For $^3$He one sees a rather good agreement between the theoretical
and experimental cross sections. Unfortunately, as mentioned above,
there are no experimental total cross section data at higher energies 
for triton.

In Fig. 3 we show the effect of the Coulomb force on the three--nucleon
photodisintegration for the case of the MT potential. 
Because of the Coulomb force one has different thresholds for $^3$He and $^3$H.
This relative shift of the two cross sections is more and more compensated
with increasing energy, and, eventually, at higher energies one finds almost
identical cross sections. Furthermore, one sees a small reduction
of the peak height due to the Coulomb force. For the
not shown case of the TN potential one finds very similar results.

Like in our previous works for the electromagnetic breakup of $^4$He
\cite{ELO97,ELO96}
we have checked the quality of the obtained cross sections with the help
of sum rules. Table 1 shows the comparison of the sum rules evaluated with the 
usual sum rule techniques with the results of an explicit integration of the 
properly weighted calculated cross sections. The
differences are about 2 \% for $\sigma_{-2}$, less than 1 \% for $\sigma_{-1}$,
and between 3 and 4 \% for $\sigma_{0}$.   
This shows the good accuracy of the obtained cross sections. From the 
rather good agreement of the $\sigma_{-2}$ sum rule one sees that the 
calculated $^3$He cross section is also reliable in the threshold region.

Here we would like to mention that the $\sigma_{-2}$ sum rule value,
\begin{equation}
 \sigma_{-2}=4\pi^2(e^2/\hbar c)\langle|\Psi_0|D_z(H-E_0)^{-1}D_z|\Psi_0\rangle, \end{equation}
was evaluated as $4\pi^2(e^2/\hbar c)\langle\Psi_0|D_z|\varphi\rangle$ 
\cite{FF84}
where $\varphi$ is the negative parity $(L=1)$ solution to the equation
\begin{equation}
(\hat{H}-E_0)\varphi=D_z\Psi_0.
\end{equation}
This sum rule is particularly interesting because it is directly connected
to the electric polarizability $\alpha_{E1}$:
\begin{equation}
\alpha_{E1} = {\hbar c \over 2 \pi^2} \sigma_{-2}(E1) \,.
\end{equation}
We obtain the following values for $^3$He: 0.143 fm$^3$ (TN potential) and  
0.151 fm$^3$ (MT potential). Therefore with our calculation we confirm the 
$\alpha_{E1}$ value of 0.15 $\pm$ 0.02 fm$^3$ extracted from photoabsorption 
experiments \cite{Faul81,Tic73,GLK91}.
For $^3$H we get 0.135 fm$^3$ (TN potential) and  0.143 fm$^3$ (MT potential)
 
Our results for $\sigma_0$ (TRK sum rule) are about 20 \% lower than 
those with realistic potentials \cite{DrK78,SchF87}. This is mainly due 
to tensor correlations,
which are neglected in our semirealistic potentials. In fact in
Ref. \cite{DrK78} it is shown that 26 \% of the TRK sum rule originates
from tensor correlations. They should affect the photodisintegration
cross sections 
mainly at higher energies with a considerable contribution beyond 
pion threshold. Thus we believe that our results should describe the
three--nucleon photodisintegration fairly well in the shown energy range.

In the following we give a summary of our work. We have calculated the 
total photodisintegration cross section of $^3$H and $^3$He with 
semirealistic potential models using the method of the Lorentz integral 
transform. We confirm the value of the polarizability of $^3$He, which
has been deduced from photoabsorption experiments. Since the used potential
models lead to a realistic $^3$H charge radius one has also a realistic
$\sigma_{-1}(^3$H) value. We find a good 
agreement of the total cross section 
with experimental data, also in the region of the peak of the dipole resonance.
This supports our prediction of a pronounced giant dipole resonance
in $^4$He, which we made in Ref. \cite{ELO96} calculating the total
$^4$He photoabsorption cross section in analogy to the present
work. Like for $^4$He it would be desirable to have better experimental
determinations of the total photoabsorption cross section for the
three--body nuclei. Also the theory can still  be improved. Effects of
multipoles others than E1, of retardation, of tensor correlations, and of 
subnuclear currents should have more and more influence at higher energies.
Also relativistic currents could have a similar effect as in deuteron
photodisintegration \cite{ArS91}.

\begin{table}
\caption{Comparison of sum rules from explicit integration (A) and 
from direct evaluation (B) with MT and TN potentials}
\begin{tabular}{|c|c|c|c|c|c|c|c|c|}
\hline
Method & A & B & A & B & A & B & A & B \\
\hline
Potential & MT & MT & MT & MT & TN & TN & TN & TN \\
\hline
Nucleus & $^3$H & $^3$H &$^3$He & $^3$He & $^3$H & $^3$H &$^3$He & $^3$He \\
\hline
$\sigma_{-2}$ [MeV$^{-1}$ mb] & 0.146 & 0.143  & 0.154 & 0.151  
& 0.138 & 0.135 & 0.147 & 0.143\\
\hline
$\sigma_{-1}$ [mb] & 2.49 & 2.48 & 2.54 & 2.53 & 2.42 & 2.42 & 2.49 & 2.47\\
\hline
$\sigma_0$ [MeV mb] & 57.8 & 56.1 & 58.1 & 55.8  & 58.6 & 57.0 & 59.0 & 56.7\\
\hline
\end{tabular}
\end{table}

\vfill\eject

\begin{figure}
\caption{Low--energy total photoabsorption cross sections for $^3$He (a) and 
$^3$H (b): theoretical results with MT and TN potentials, experimental results
with error range (dotted curves) from Ref. [1]. In (a) also the sum of the 
experimental two-- and three--body cross sections from Ref. [21] is shown 
(filled circles). The total error is assumed to be the square root of the sum 
of the squared errors.}
\end{figure} 

\begin{figure}
\caption{As Fig. 1, but for an extended energy range up to pion threshold}
\end{figure} 

\begin{figure}
\caption{Cross sections
for $^3$H (dashed curve) and $^3$He (full curve) with MT potential} 
\end{figure} 

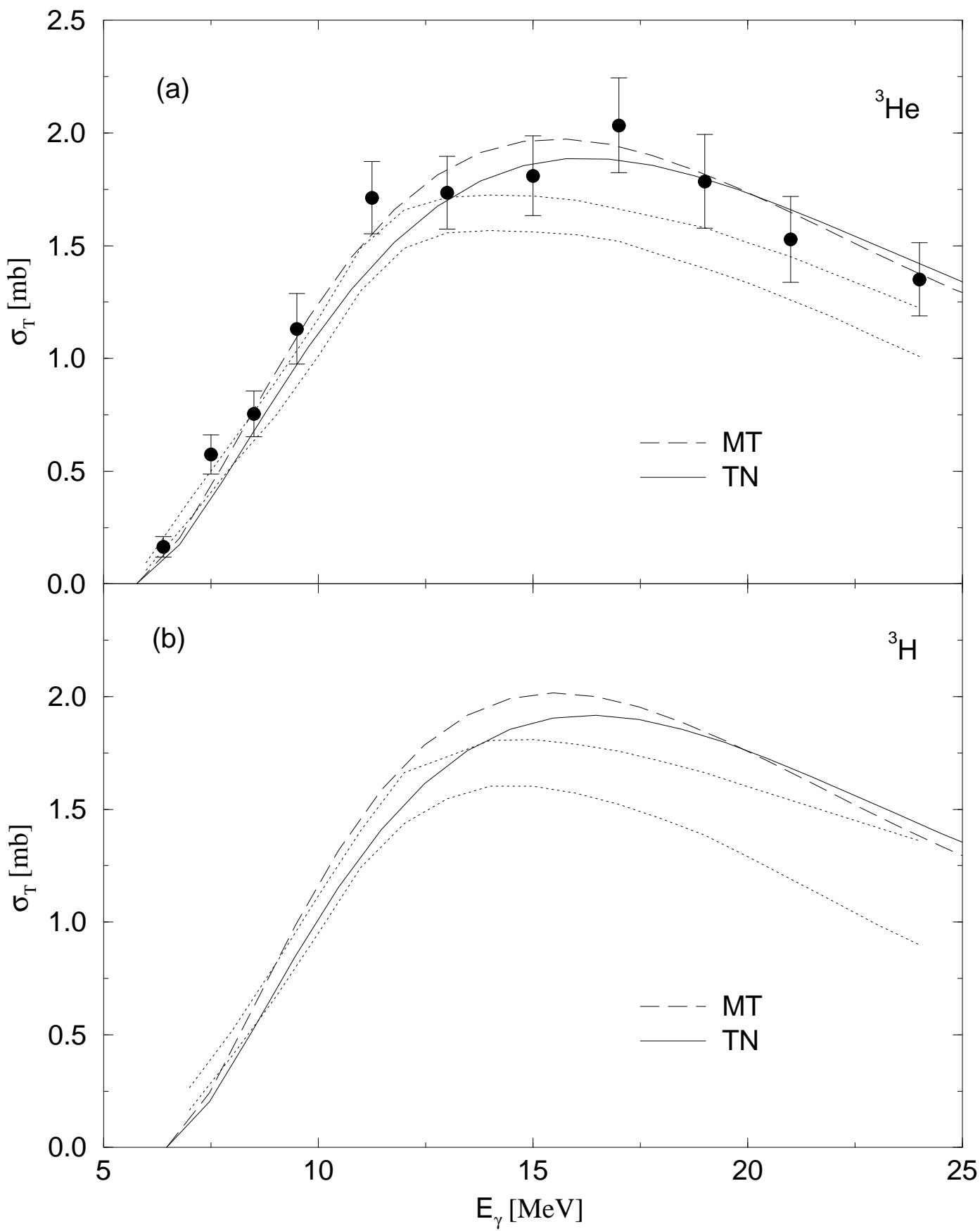
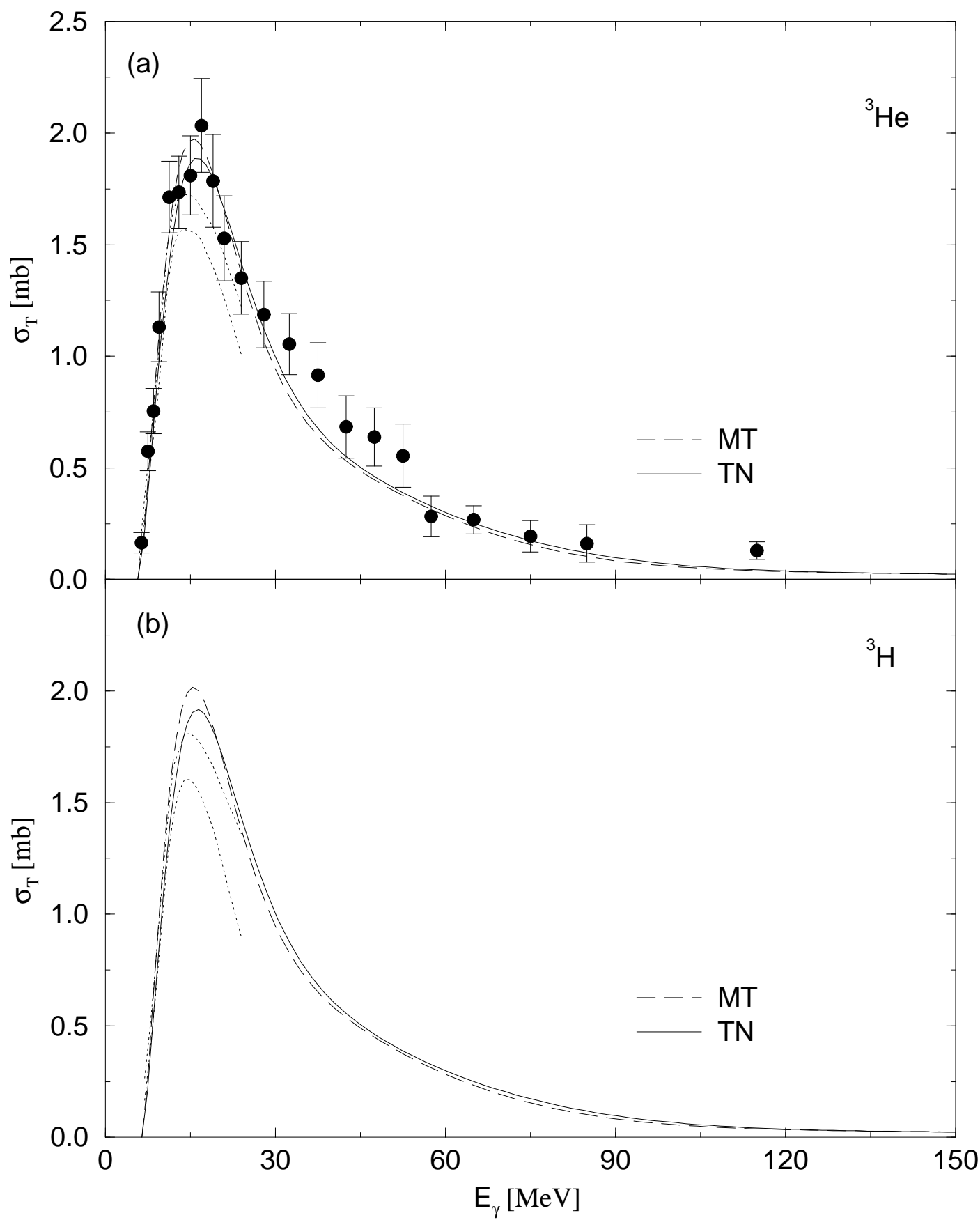
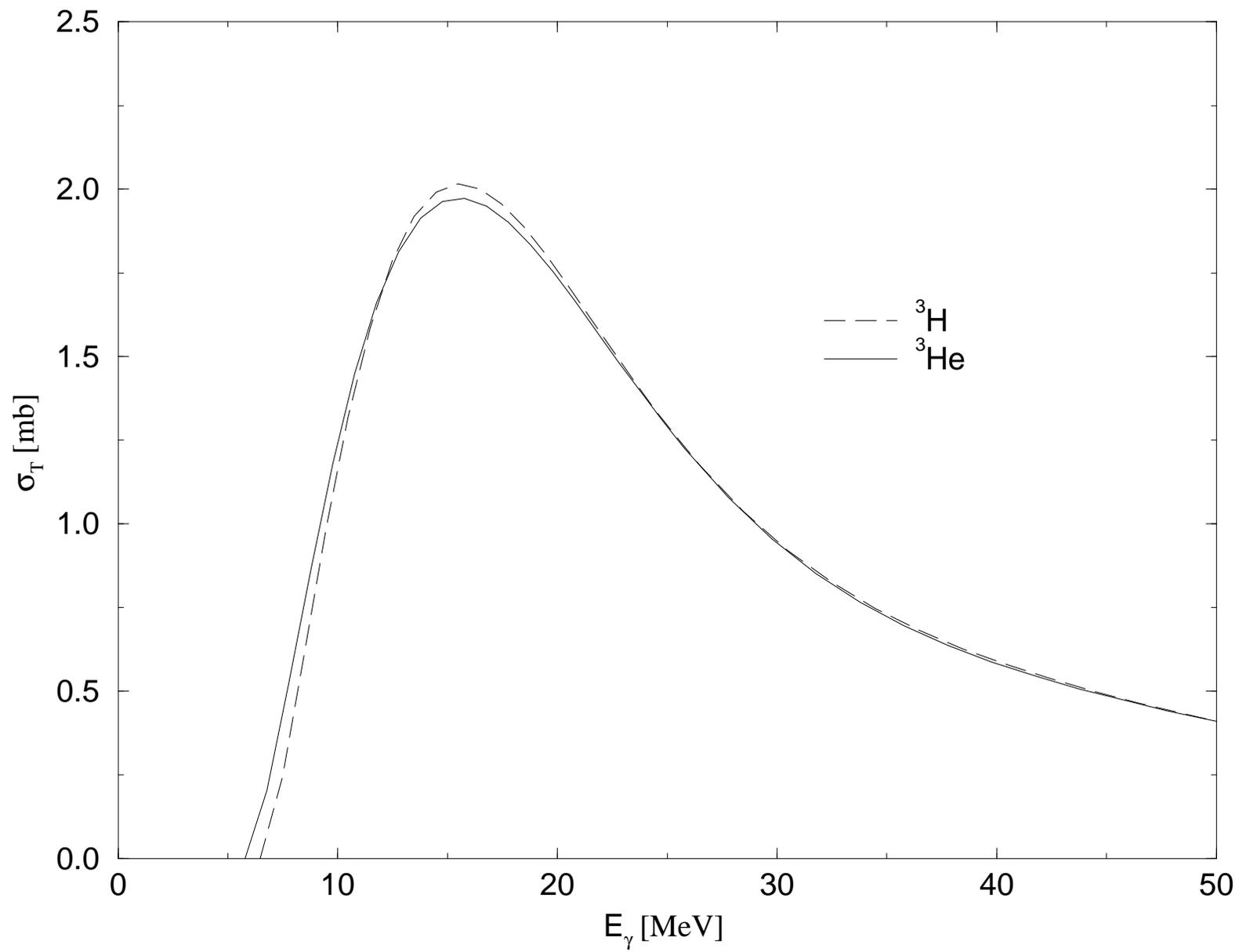
\end{document}